\documentstyle[multicol,aps,prl,epsf,psfig]{revtex}
\begin{document}
\def\be{\begin{equation}}
\def\ee{\end{equation}}

\title{Novel Position-Space Renormalization Group \\
for Bond Directed Percolation in Two Dimensions}
\author{H\"useyin Kaya$^{a}$ and Ay\c se Erzan$^{a,b}$}
\address{$^a$TUBITAK, Feza Gursey Institute, P.O. Box 6 \\
\c Cengelk\"oy 81220, Istanbul - TURKEY}
\address{$^b$Istanbul Technical University, Faculty of Science and
Letters \\
Department of Physics, Maslak 80626, Istanbul-TURKEY}
\maketitle
\begin{abstract}
A new position-space renormalization group approach is
investigated for bond
directed percolation in two dimensions. The threshold value 
for the bond occupation probabilities is
found to be $p_c=0.6443$. Correlation length exponents on time
(parallel)
and space (transverse) directions are found to be
$\nu_{\parallel}=1.719$ and $\nu_{\perp}=1.076$, respectively,
which are in
very good agreement with the best known series expansion
results.
\end{abstract}

\begin{multicols}{2}
\section{Introduction}

Directed percolation has been studied extensively [1-6],
since
it plays very important role in a large
variety of non-equilibrium systems with a single absorbing state
\cite{pgrass}.
A wide array of dynamic processes as fluid flow
through a porous medium in a external field \cite{flux},
forest fires \cite{ffm1,ffm2} or epidemic growth models \cite{epidemic},
reaction-diffusion systems \cite{zgb,schlogl,grinstein},
damage spreading \cite{grass},
self-organized criticality \cite{bak},
models of growing surfaces with roughening transition
\cite{tang,tang1,olami}, etc.
fall into  the same universality class as directed percolation.
There is no exact result for the critical exponents
characterising the dynamical phase transition seperating the
absorbing steady
state from the active phase. The correlation length exponents in the
longitudinal 
and transverse directions, defined via $\xi_{\parallel} \sim |p-
p_c|^{-\nu_\parallel}$ and
$\xi_{\perp}\sim |p-p_c|^{-\nu_\perp}$ 
 and the percolation threshold  $p_c$ have 
been  
found to great accuracy  by 
series expansion methods~\cite{series1,series2} 
to be $\nu_\parallel =1.733$,                
$\nu_\perp =1.097$ and $p_c=0.6447$, respectively.

The fixed-scale transformation (FST) \cite{erzan2} introduced by
Pietronero, Erzan and Evertsz~\cite{erzan,erzana} has been used to
investigate the
self-similar properties of irreversible growth models, namely,
diffusion limited aggregation~\cite{witten} and dielectric
breakdown~\cite{niemeyer}, 
as well as  other
intrinsically critical growth problems such as cluster-cluster
aggregation~\cite{nero1}. 
The FST has also been applied to systems such as
percolation~\cite{nero0},
Ising or Potts models~\cite{erzan1}, invasion percolation
\cite{nero,piet1}
and to
directed percolation~\cite{erzan3} at the critical 
point.
Renormalization group ideas have been used in conjunction with
the FST approach~\cite{FSTRG}, in order to identify the scale
invariant growth rules in terms of which the finite cell fixed
scale transformation matrix should be computed, with good
results.  Forest fires~\cite{ffires} (which fall into the same
universality class as directed percolation) and the sandpile
model~\cite{FSTsoc,vespi1} have been studied by means of the
``dynamically driven renormalization group''~\cite{DynRG}, which
combines  real space renormalization group ideas with a dynamical
steady state condition reminiscent of the fixed scale
transformation approach.

In this paper, we would like to introduce a novel position space
renormalization group (PSRG) treatment of the directed
percolation problem.  This approach modifies conventional PSRG
methods in two ways:  {\it i) } the weights of different initial
states in the RG cell are computed from the steady state
distribution found from the  fixed point of the FST; 
{\it ii)} a ``dynamical" coarse graining procedure is defined 
which allows
for the appearance of two different scale factors in the
longitudinal (time-like) and transverse (space-like) directions,
thus taking into account the self-affine nature of the problem. 
These scale factors are determined independently, without having
to make any additional assumptions.
 
The paper is organized as follows.  In the next section, we
construct the renormalization group transformation, subject to
the dynamical fixed point condition, and compute the critical
parameter as well as the transverse and longitudinal correlation
length exponents.  In the last section we provide a short
discussion.

\section{RG Transformation with Dynamical Coarse Graining}

We consider 1-time and 1-space dimensional bond directed
percolation as a
growth process. Let us recall the growth rules: if a site at time
$t$ is active, its two neighbours
at time $t+1$ may be activated,  each independently, with a
probability $p$.
If $p$ is larger than the threshold value $p_c$, there is a finite
probability
that the growth process is continued indefinitely.

We now outline a renormalization group procedure
which takes into account fluctuations in regions larger
than the renormalization group cell by using a steady-state
distribution of initial conditions, obtained from the FST. It also
introduces self-consistently determined rescaling
lengths, rather than pre-set scale factors between the original
and coarse grained lattices.  

Our renormalization group cell is shown in Fig. 1(a). The boxes
$AA^\prime BB^\prime $ and $CC^\prime A^{\prime}A^{\prime\prime }$
will coarse
grain, under a dynamic coarse graining procedure we describe
below, to the bonds $ab$ and $ac$, as shown in 
Fig.~1(b).  This process conserves the transverse and
longitudinal directions, but does not conserve the lattice angles, 
since
these two directions are scaled differently. Rescaling by the
appropriate longitudinal and transverse scale factors will
restore the original lattice (Fig. 1(c)). The lattice spacing is
taken to be $\sqrt{2}$ for convenience, yielding $\ell_\perp=1$
and $\ell_\parallel =1$ for the transverse and perpendicular
distances between the nearest neighbors, such as $AA^\prime $ and
$BB^\prime $.  Clearly, we only have to consider one of the boxes,
e.g., $AA^\prime BB^\prime $ for our renormalization procedure, by
symmetry.
\begin{figure}
\begin{center}
\leavevmode
\psfig{figure=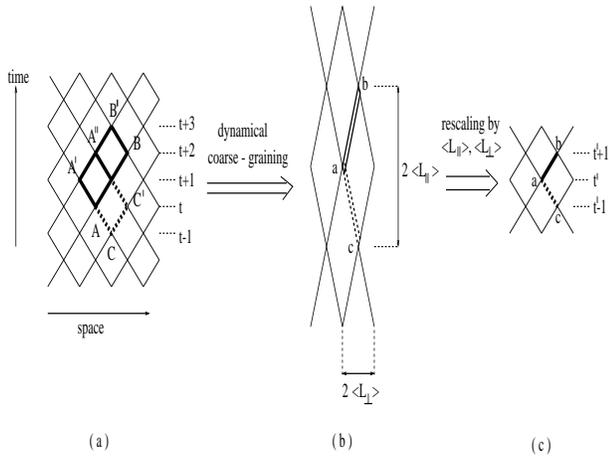,width=8cm,height=6cm,angle=0}
\end{center}
\narrowtext
\caption{The dynamical coarse-graining procedure. $(a)$ The RG cells
$AA'BB'$ and $CC'A'A^{\prime\prime}$ (the bold lines) in the original
lattice are coarse-grained $(b)$ to the bonds $ab$ and $ac$ respectively,
which are rescaled in the next step $(c)$ to preserve the lattice angles.}
\end{figure}  
To obtain the renormalization transformation for the bond
occupation probability $p$, we compute the total probability
 $P(p)$ that a spanning path exists across the box
$AA^\prime BB^\prime $, by considering all the different initial
configurations (states of $A$ and $A^\prime $) and the spanning
configurations that can be obtained from them.  A spanning
configuration is defined as one where a path starts from either
$A$, or $A^\prime $ or both, and ends on $B$ or $B^\prime $ or
both.

\subsection{Steady state distribution of initial conditions}

Since there are two possible origins which can be active both
together or by
themselves, four different initial configurations have to be 
considered as
illustrated in Fig.2.
Any spanning configuration may have more than one path which
connects the origins to the end points.  For instance, two
different paths are possible for connecting the sites $A$ and
$B^{\prime }$ in the spanning configuration shown in Fig.3.
\begin{figure}
\begin{center}
\leavevmode
\psfig{figure=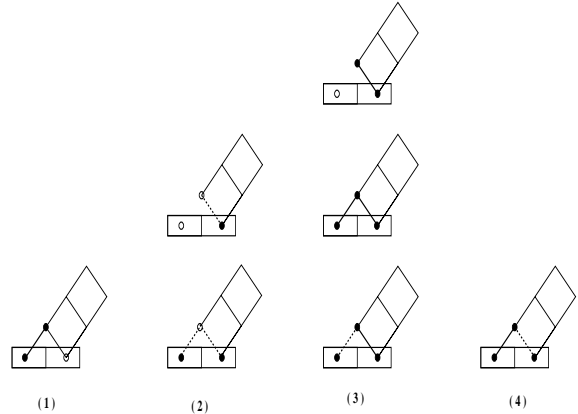,width=8cm,height=6cm,angle=0}
\end{center}
\narrowtext
\caption{ Four different initial configurations are represented. In
computing their respective weights, (see text) the bold lines in
the each configuration are replaced with the probability $p$, and
the dashed lines with $1-p$.}
\end{figure}

In order to compute the probabilities of finding each initial
configuration,
say $W_i$, $i=1,...,4$, we make use of the time invariant
probabilities for the relative frequency of doubly or singly
occupied boxes in a box covering of the equal-time transverse
subsets on the cluster of active sites, in terms of the
probabilities $p$.  
\begin{figure}
\begin{center}
\leavevmode
\psfig{figure=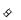,width=8cm,height=6cm,angle=0}
\end{center}
\narrowtext
\caption{ Two different paths belonging to the same spanning
configuration.}
\end{figure}

In recent work~\cite{erzan3}, the FST approach has been used to
calculate the fractal dimension of directed percolation at the
critical point, given the  threshold
value of $p$. The main idea underlying the FST approach here is
that at the critical point, as  $t\rightarrow \infty$, the
infinite transverse subsets  of active sites at any given $t$ are
statistically similiar under translation in $t$.   In particular,
they can be modeled by Generalized Cantor Sets generated by a
random sequence of
fragmentations obeying a one-parameter, scale invariant
distribution as
shown in Fig.4. Given a hierarchical partitioning of the
transverse subsets into cells of size
$2^{-k}$, the relative probability of encountering singly
occupied or doubly occupied  cell among nonempty cells at a 
fixed arbirary
scale is equal to $C$ or $1-C$. This probability can be
computed~\cite{erzan3} from the fixed point of the fixed-scale 
transformation as,
\be
{C={\frac{2-3p+5p^2-\sqrt{36-108p+109p^2-30p^3+9p^4}}{{2p^2+4p
-4}}}}\;\;,\ee
up to first order in the FST approach.
\begin{figure}
\begin{center}
\leavevmode
\psfig{figure=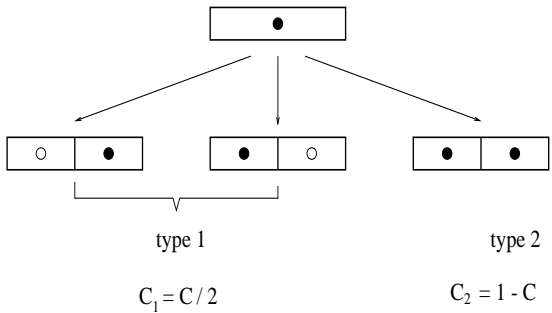,width=8cm,height=6cm,angle=0}
\end{center}
\narrowtext
\caption{ The generator for the random Cantor set corresponding to a
transverse subset of the incipient infinite cluster. Cells of
type-1 and type-2 participate in the fine graining process with
corresponding probabilities
$C_1$ and $C_2$ . We represent the activated sites with black
circle and inactive sites with
white circle.}
\end{figure}

We now make use of this FST results to compute the normalized weights $W_i$
of the initial configurations $i$
in terms of $C$ and $p$, 
\begin{eqnarray}
W_1\, &=&\, p\, C\, /\, 2\\
W_2\, &=&\, (1-p)\, C\, /\, 2\, +\, (1-p)^2\, (1-C)\\
W_3\, &=&\, p\, (1-C/2)\\
W_4\, &=&\, p\, (1-p)\,\, (1-C)\,\, .\end{eqnarray}

In the following subsection we proceed to obtain the renormalization
transformation in terms of which the critical value of $p$ can be determined.

\subsection{The renormalization transformation for $p$}

Depending on the initial configurations $i=1,...,4$, the total number of
possible
spanning configurations will be different. One sees that in
Fig.2 $(1)$
only one; $(2)$ seven; and $(3)-(4)$ eighteen spanning
configurations each are
possible. It should also be noticed that some spanning
configurations can be
observed in more than one different initial configuration.
For example, the spanning configuration which contains only one path
which starts from 
$A^{\prime }$ and ends at $B^{\prime }$ can be observed in all the
initial
configurations except the second one (see Fig.2). The total probability
$f_i$
of the
spanning cluster for the $i$'th initial configuration is given
by 
\begin{eqnarray} f_1(p) &=&p^2\\
f_2(p)&=&p^2+p^3-p^4\\
f_3(p)&=&2p^2+p^3-3p^4+p^5\\
f_4(p)&=&f_3(p)\;\;.\end{eqnarray}
The renormalization group transformation for $p$ is then found
to be 
\be P(p)=W_1f_1(p)+W_2f_2(p)+W_3f_3(p)+W_4f_4(p)=p^{\prime }\,\,\,
.\ee
The fixed point of this transformation 
gives the threshold value, $p_c = 0.6443$ which is in 
agreement up to the third digit with the series expansion \cite{series1,series2}
result, namely $0.6447$.

\subsection{The Affine transformation in the longitudinal and
transverse directions}

The system has two independent correlation lengths
$\xi_{\parallel}$ and
$\xi_{\perp}$, parallel and perpendicular to the time direction
respectively,
which diverge with different exponents as $\xi_{\parallel}\sim
|p-p_c|^{-\nu_{\parallel}}$ and $\xi_{\perp}\sim
|p-p_c|^{-\nu_{\perp}}$. To
compute these exponents, we use the 
eigenvalue equations,
\be{dp^{\prime }\over
{dp}}\Big|_{p=p_c}=b_{\parallel}^{1/\nu_{\parallel}},
\,\,\,\,\,\,\,\,\,\,\,\,\,\,\,\,\,\,
{dp^{\prime }\over {dp}}\Big|_{p=p_c}=b_{\perp}^{1/\nu_{\perp}}\,
.\ee
We determine  the appropriate rescaling lengths $b_\parallel $
and $b_\perp$ from,
\be b_\parallel = {<L_\parallel>\over l_\parallel},
\,\,\,\,\,\,\,\,\,\,\,\,
b_\perp\, =\, {<L_\perp>\over l_\perp}\, ,\ee
in terms of the average projected lengths,  $<L_\parallel >$ and
$<L_\perp >$, of the spanning paths onto  the time and
transverse directions as shown in Fig.5. These quantities are the amounts by
which the  coarse grained lattice (see Fig.1(b)) has been dilated in
the ``dynamical coarse-graining'' step.          
Since we have taken the lattice constant to be  $\sqrt 2$,  the
projections of
a single bond on the original lattice onto the time and space
directions are $\ell_\perp=\ell_\parallel =1$.
\begin{figure}
\begin{center}
\leavevmode
\psfig{figure=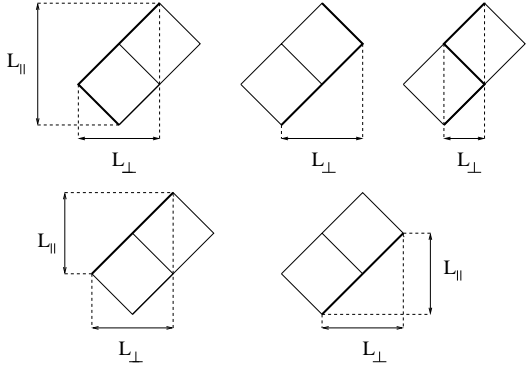,width=8cm,height=6cm,angle=0}
\end{center}
\narrowtext
\caption{All possible paths and their extremal
projections contributing to the dynamical rescaling factors
$<L_\parallel>$ and $<L_\perp>$. See Fig.1.}
\end{figure}

In the time direction, one has to take into account the fact
that under coarse graining, different time steps collapse onto
each other; that is why we consider each path originating from
$A$ or $A^\prime $ and terminating on $B$ or $B^\prime $ as
contributing seperately to $L_\parallel$. While enumerating the
possible paths over which the
average is taken, it makes a difference if there is a bond
between $A$ and $A^{\prime }$, since the paths change in case 
$A^{\prime }$ is
activated by $A$. The same thing holds for the end sites $B$ and
$B^{\prime }$. Note that the result of taking the extremal projections first
and then averaging is different from finding the ``average path'' and then
taking its projections. This point will be further discussed below.

For $X \equiv\{ {\perp, \parallel}\}$, we have
\be<L_{X}>=\sum_iW_i\sum_\beta q_{i,\beta} \sum_\alpha
L_{X}^{i,\beta ,\alpha}\ee
where $q_{i,\beta}$ is the relative probability of any spanning
path $\beta$ belonging to an initial state $i$, and is to be found
from $f_i$ by giving equal weights to the distinct paths in any spanning
configuration
equally.
Substituting the value of $p_c$ found from Eq.(10) into Eqs.(2,3), we find, 
\be <L_\perp> =1.7996\ee
and
\be <L_\parallel>=2.5561\,\, .\ee

These values yield, together with Eq.(11), 
the correlation length exponents  to be
$\nu_\parallel =1.719$ and $\nu_\perp =1.076$ which are
comparable with the
best known results \cite{series1,series2} $\nu_\parallel =1.733$
and
$\nu_\perp =1.097$.

\section{Discussion}

There have been some earlier studies of directed (therefore self-affine)
systems via position space renormalization group (PSRG) techniques.
Introduction of two different scaling factors $b_\parallel$ and $b_\perp$
due to existence of two independent scaling directions was firstly suggested
by Dhar and Phani \cite{dhar}, where they employed a decimation
transformation.  However, their results are far from the accepted values of
the critical exponents and dependent upon their choice of the RG cell, which
determines $b_\parallel /b_\perp$. Very large-cell PSRG calculations by Zhang
and Yang \cite{yang} are able to accurately reproduce the self-affine
behaviour of directed self-avoiding walks, but are more appropiately in a
class with Monte Carlo RG. A bond-moving and decimation transformation for
anisotropic directed bond percolation in arbitrary dimension \cite{droz1} and
its generalization for other directed systems \cite{droz2} by da Silva and
Droz give the critical fugacity very accurately in all dimensions for directed
self-avoiding-walks. For two dimensional directed percolation, it yields good
results for the threshold value $p_c$ and $\nu_\parallel$, but is not as good
for $\nu_\perp$. The approach in these papers is in fact very close to ours;
however, the way the projections of paths are taken to compute the scaling
factors in the two different directions are different.

One may consider two different ways of projecting a path. The
approach of da Silva and Droz \cite{droz1,droz2} is to draw
a vector between the origin and the end point, and
take its 
two components to  be the projections of the path. Our is to
take the projection to be the transverse or longitudinal distance
between the {\it extremal} points of the  path, as shown in Fig.5.
\begin{figure}
\begin{center}
\leavevmode
\psfig{figure=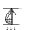,width=8cm,height=6cm,angle=0}
\end{center}
\narrowtext
\caption{ Macroscopic paths with identical end-to-end vectors, but
strongly differing extremal projections.}
\end{figure}

We believe that our method 
yields a result which corresponds
more closely to what is meant by the size of the  
transverse and longitudinal fluctuations, in that it measures
more accurately the actual size of the region over which a
coherent flow takes place.
This becomes more evident if we consider 
the incipient infinite cluster of our system for $p\sim p_c$. In
this cluster, there can be many paths which connect some point
at time $t_0$ 
to an  end point at $t_0+t$. Even though the
total number of bonds in these different paths are the same, the
spatial size of the spanned
regions can be radically different (see Fig. 6). Since the
transverse (longitudinal) correlation length  
corresponds to the  spatial (time-like) size of the fluctuations,
it is more appropriate  to take the average over the extremal
transverse (longitudinal) extent of each path.  

In conclusion, by incorporating the FST fixed point condition in the
determination of our distribution of initial configurations, and by a new,
dynamical coarse graining procedure which makes use of averages over
extremal projections of spanning paths, we have succeeded in computing the
percolation threshold and the correlation function exponents much more
accurately than before.

{\it Note added in proof:} D. Sornette has kindly brought to our attention
Refs.\cite{sornet1,sornet2} where the equivalence of directed percolation
to a parallel version of the Bak-Sneppen model~\cite{baka} has been
rigorously demonstrated. Also see~\cite{vespi} for a comparison of
directed percolation with sandpile models.


\end{multicols}
\end{document}